\documentclass[runningheads]{llncs}
\usepackage{graphicx}
\usepackage{subcaption}
\usepackage{xcolor}
\usepackage{indentfirst}
\begin{document}
\title{VertXNet: Automatic Segmentation and Identification of Lumbar and Cervical Vertebrae from Spinal X-ray Images}
\titlerunning{VertXNet: Vertebrae Segmentation and Identification of Spinal X-Ray}
\author{
Yao Chen $^{\ast}$ \inst{1} 
Yuanhan Mo $^{\ast}$\inst{2}
Aimee Readie\inst{1}
Gregory Ligozio\inst{1}\\
Thibaud Coroller $^{\dagger}$ \inst{1} 
Bart\l omiej W. Papie\.z $^{\dagger}$ \inst{2}
}
\authorrunning{Y. Chen et al.}
\institute{Novartis Pharmaceutical Company, East Hanover, NJ, USA \\
\email{\{yao.chen, aimee.readie, gregory.ligozio, thibaud.coroller\}@novartis.com} \and
Big Data Institute, University of Oxford, Oxford, UK\\
\email{yuanhan.mo@ndm.ox.ac.uk, bartlomiej.papiez@bdi.ox.ac.uk}\\
}

\maketitle              % typeset the header of the contribution
\def\thefootnote{$\ast$}\footnotetext{Contribute equally}

\def\thefootnote{$\dagger$}\footnotetext{Contribute equally}

\begin{abstract}
Manual annotation of vertebrae on spinal X-ray imaging is costly and time-consuming due to bone shape complexity and image quality variations.
In this study, we address this challenge by proposing an ensemble method called VertXNet, to automatically segment and label vertebrae in X-ray spinal images.
VertXNet combines two state-of-the-art segmentation models, namely U-Net and Mask R-CNN to improve vertebrae segmentation. 
A main feature of VertXNet is to also infer vertebrae labels thanks to its Mask R-CNN component (trained to detect 'reference' vertebrae) on a given spinal X-ray image.
VertXNet was evaluated on an in-house dataset of lateral cervical and lumbar X-ray imaging for ankylosing spondylitis (AS) patients.
Our results show that VertXNet can accurately label spinal X-rays (mean Dice of 0.9). It can be used to circumvent the lack of annotated vertebrae without requiring human expert review. This step is crucial to investigate clinical associations by solving the lack of segmentation, common bottleneck for most computational imaging projects. 

\end{abstract}
\section{Introduction}
Reliable vertebrae segmentation and identification from spinal X-ray images is a prerequisite to quantitatively perform functional analysis of the spine.
% and also is the fundamental step to prove the clinical association between imaging data and the clinical endpoints.
However, this step is challenging due to the need for large datasets to efficiently train machine learning models whilst relying on human annotation of the images. 
Using clinical experts to annotate images is time-consuming and expensive, but also prone to errors (i.e. inter- and intra-reader variability). 
Therefore, a solution that could automatically segment and label images would greatly decrease the cost, time and inter-observer errors caused by human experts.
% to reducing both manual labour, and inter-observer variability of the outputs created. 
In this paper, we propose a novel method,VertXNet, for vertebrae segmentation and identification from spinal X-ray images that was validated utilizing anonymized internal trial data (MEASURE1) with ankylosing spondylitis (AS) patients \cite{baeten2015secukinumab}.

\begin{figure}[!t]
  \includegraphics[width=0.95\linewidth]{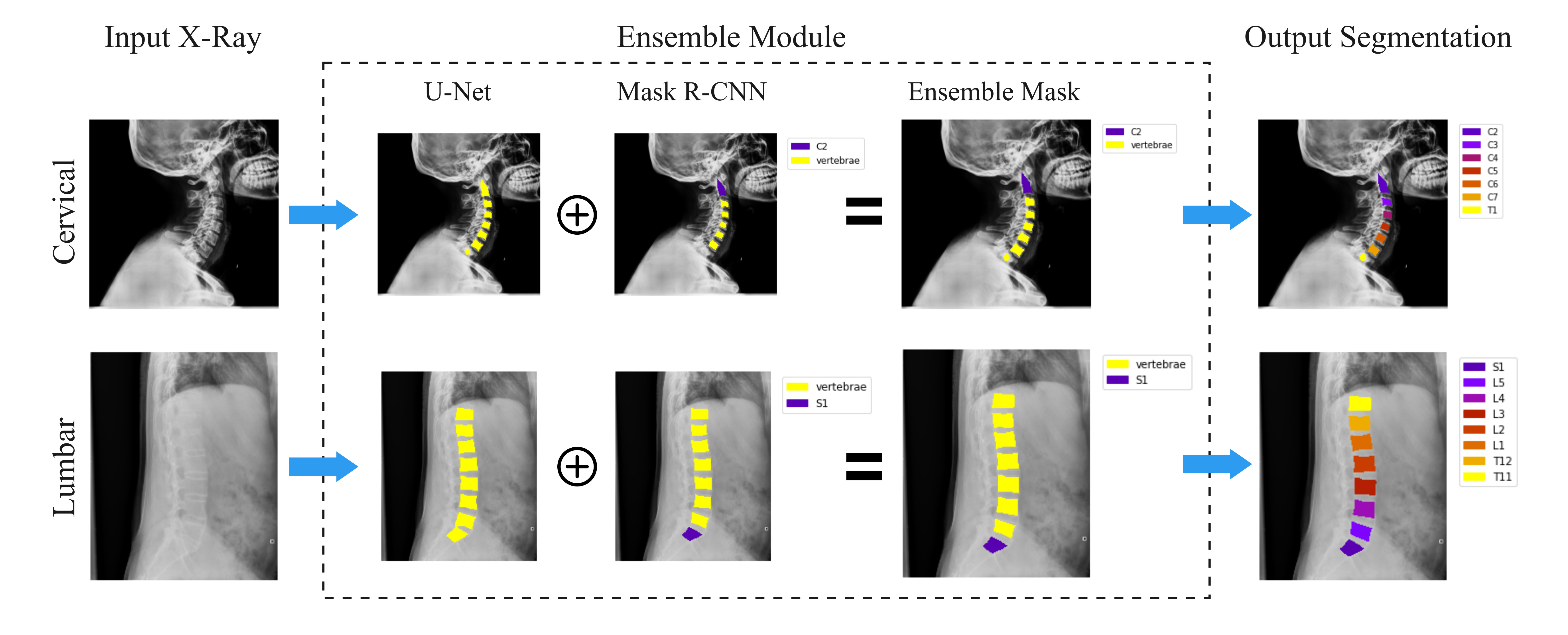}
  \caption{VertXNet overview: 1) An ensemble method (U-Net and Mask R-CNN) produces a robust segmentation for each vertebra, 2) Mask R-CNN locates a reference vertebra (either C2 or S1) and infers remaining vertebrae.}
  \label{fig:pipeline}
\end{figure}

\section{Method}
% VertXNet employs the state-of-the-art (SOTA) segmentation models, namely U-Net \cite{ronneberger2015u} and Mask R-CNN \cite{he2017mask}, to produce all possible contours for each vertebrae appearing in a cervical or lumbar X-ray. 
% We decided to combine the outputs from the two models to generate more robust segmentation. 
% The ensemble mask not only solves the potential risk of overlapping masks or missing vertebrae masks by one model, but also improve the Dice coefficients of the outputs.
% The overview of our method (VertXNet) is shown in Figure \ref{fig:pipeline}.

VertXNet employs two-step pipeline, segmentation and identification of spinal vertebrae (see Fig. \ref{fig:pipeline}).
X-ray images are input to a segmentation module to first generate separable mask of each vertebra. Two state-of-the-art methods, U-Net \cite{he2017mask} and Mask R-CNN \cite{ronneberger2015u}, have been investigated; however, neither of them works perfectly individually. U-Net generates connected masks when two neighbouring vertebrae are too close. Mask R-CNN misses vertebrae if they are incomplete or have unclear boundaries. A rule-based ensemble method is then proposed to combine outputs of both methods to produce complete vertebrae segmentation.

Due to the similarity in shape, it is extremely challenging to correctly label the vertebrae under multi-class task settings (see Table \ref{tab:metrics}). The spatial relationship among vertebrae, such as the knowledge that L1 should be followed by L2, can be used. Thus
finding the reference vertebrae becomes essential to properly label all remaining vertebrae in an X-ray image. 
Luckily, both lumbar and cervical images contain two types of vertebrae that can be easily distinguished from others, and are identified as the reference vertebrae.
The reference vertebrae include: 1) ‘C2’, cone-shaped and the first detectable vertebra at the top of the cervical X-ray image, and 2) ‘S1’, a triangular-shaped vertebra, which is commonly the last visible vertebra at the bottom of a lumbar X-ray. To detect the reference vertebrae, the Mask R-CNN is trained to distinguish the reference vertebra from other vertebrae.
If C2 is detected on an image, we simply "zip" down the spine to infer the cervical (C) and thoracic (T) labels (from C3 to C7 and T1). 
On the other hand, if "S1" is detected we "zip" up the spinal vertebrae to infer the lumbar (L) and thoracic (T) labels (from L5 to L1, and T12 to T11).

% and once the reference vertebrae are located by the model, the labelling for other vertebrae is then trivial.

\section{Experiments and Results}
{\bf Data.} The method was developed based on the anonymized dataset of a secukinumab clinical trial for Ankylosing Spondylitis. 
A sample of images from the MEASURE~1 CAIN457F2305 study \cite{baeten2015secukinumab} were utilized, in which sagittal cervical and lumbar X-ray images were acquired at different visit (Baseline, Week 104, and Week 208).
All vertebra in the 512 X-ray images (293 cervical and 219 lumbar) were annotated by an expert radiologist. 

{\bf Experiments.} The 512 annotated X-rays were randomly split and stratified by spinal acquisition type. 80\% of the X-rays images were used for training the pipeline and the remaining 20\% for testing. 
All models were trained on the training dataset and the Dice coefficients were calculated on the testing set to compare all models. 
Our method's mean Dice coefficient was {\bf 0.90}, compared with Mask R-CNN's {\bf 0.73} and U-Net's {\bf 0.72}. Detailed Dice coefficients on each vertebral type have been provided in Table \ref{tab:metrics}. 
Drop in performances across models were observed for T1, T11 and T12 due to data imbalance caused by the lack of appearance of these vertebrae in X-rays. 
Greater performance from our model was observed for T1, T11 and T12.
In the experiment, the ensemble mask not only solved the potential risk of overlapping masks or missing vertebral masks by one model, but also improved the Dice coefficients of the outputs.

\begin{table}[!htbp]
    \caption{Model Comparison}
    \begin{subtable}[h]{0.9\textwidth}
        \centering
        \begin{tabular}{lccccccc}
            Model & C2 & C3 & C4 & C5 & C6 & C7 & T1 \\
            \hline \hline
            Mask RCNN & 0.852 & 0.882 & 0.889 & 0.842 & 0.767 & 0.736 & 0.325 \\
            \hline
            U-Net & 0.823 & 0.846 & 0.848 & 0.822 & 0.809 & 0.717 & 0.067 \\
            \hline
            % Our method & \bf 0.8878 & \bf 0.8952 & \bf 0.8968 & \bf 0.9069 & \bf 0.8927 & \bf 0.8451 & \bf 0.5936 \\
            VertXNet & \bf 0.905 & \bf 0.912 & \bf 0.912 & \bf 0.906 & \bf 0.901 & \bf 0.881 & \bf 0.518 \\
            \hline
        \end{tabular}
        \caption{Dice coefficients of Cervical vertebrae on test set}
    \end{subtable}
    \begin{subtable}[h]{0.9\textwidth}
        \centering
        \begin{tabular}{lcccccccc}
            Model & T11 & T12 & L1 & L2 & L3 & L4 & L5 & S1 \\
            \hline\hline
            Mask RCNN & 0.379 & 0.577 & 0.706 & 0.730 & 0.615 & 0.707 & 0.732 & 0.744 \\
            \hline
            U-Net & 0.338 & 0.519 & 0.643 & 0.726 & 0.785 & 0.772 & 0.726 & 0.603 \\
            \hline
            % Our method & \bf 0.6786 & \bf 0.7303 & \bf 0.8112 & \bf 0.8210 & \bf 0.8256 & \bf 0.8136 & \bf 0.8333 & \bf 0.7862 \\
            VertXNet & \bf 0.761 & \bf 0.843 & \bf 0.864 & \bf 0.870 & \bf 0.871 & \bf 0.865 & \bf 0.854 & \bf 0.823 \\
        \end{tabular}
        \caption{Dice coefficients of Lumbar vertebrae on test set}
    \end{subtable}
    \label{tab:metrics}
\end{table}

% \subsection{Conclusion}
{\bf Conclusion.} We successfully demonstrated that our pipeline can automatically segment and label vertebrae from spinal X-ray images and has outperformed two benchmark models for the segmentation task according to the Dice coefficients.
The proposed method is the fundamental step to perform the further quantitative analyses.

\newpage
\bibliographystyle{splncs04}
\bibliography{main}




% LNCS Homepage, \url{http://www.springer.com/lncs}. Last accessed 4
% \end{thebibliography}
\end{document}